\documentclass[%
 aip, apl,
 amsmath,amssymb,
 reprint,%
]{revtex4-1}

\usepackage{graphicx}
\usepackage{dcolumn}
\usepackage{bm}

\usepackage[utf8]{inputenc}
\usepackage[T1]{fontenc}
\usepackage{mathptmx}
\usepackage{etoolbox}
\usepackage{siunitx}
\usepackage{epsfig}
\usepackage{epstopdf}
\usepackage{siunitx}
\usepackage{textcomp}
\usepackage{color, soul}
\usepackage{dcolumn}
\usepackage{siunitx}
\usepackage{xcolor}
\usepackage{tcolorbox}
\tcbuselibrary{skins}
\newtcbox{\inlinehl}{on line,
  boxsep=0pt,
  left=1pt,
  right=1pt,
  top=0pt,
  bottom=0pt,
  colback=yellow,
  colframe=yellow,
  arc=2pt,
  boxrule=0pt
}

\usepackage{array} 

\newcolumntype{L}{>{$}l<{$}}
\newcolumntype{R}{>{$}r<{$}}

\newcolumntype{d}[1]{>{\ensuremath}D{#1}{#1}{-1}}

\newcommand{\ii}{{\rm i}}
\newcommand{\micm}{{\rm \mu m}}
\newcommand{\micl}{{\rm \mu L}}

\makeatletter
\def\@email#1#2{%
 \endgroup
 \patchcmd{\titleblock@produce}
  {\frontmatter@RRAPformat}
  {\frontmatter@RRAPformat{\produce@RRAP{*#1\href{mailto:#2}{#2}}}\frontmatter@RRAPformat}
  {}{}
}%
\makeatother
\begin{document}

\preprint{AIP/123-QED}

\title[Mic drop]{'Mic drop': on estimating the size of sub-mm droplets using a simple condenser microphone}

\author{Avshalom Offner}
\affiliation{ 
School of Mathematics and Maxwell Institute for Mathematical Sciences, The University of Edinburgh, Edinburgh EH9 3FD, UK}
\email{avshalom.offner@ed.ac.uk}

\date{\today}

\begin{abstract}
The size distribution of aerosol droplets is a key parameter in a myriad of processes, and it is typically measured with optical aids (e.g., lasers or cameras) that require sophisticated calibration, thus making the measurement cost intensive. We developed a new method to indirectly measure the size of small droplets using off-the-shelf <\$1 electret microphones. In this method we exploit the natural oscillations that small droplets undergo after impacting a flat surface: by allowing droplets to land directly on a microphone diaphragm, we record the impact force they exert onto it and calculate the complex resonant frequencies of oscillations, from which their size can be inferred. To test this method, we recorded the impact signals of droplets of varying sizes generated by a pipette and extracted the resonant frequencies that characterize each signal. Various sources of uncertainty in the experiments led to a range of frequencies that can characterize each droplet size, and hence a data-driven approach was taken to estimate the size from each set of measured frequencies. We employed a simple setting of neural network and trained it on the frequencies we measured from impact of droplets of prescribed radius. The network was then able to predict the droplet radius in the test group with an average error of 2.7\% and a maximum of 8.6\% relative to the pipette nominal values. These results, achieved with a data set of only 320 measurements, demonstrate the potential for reliable size-distribution measurements via a simple and inexpensive method. 
\end{abstract}

\maketitle

Aerosols that are composed of liquid droplets drive a variety of natural and industrial processes, including the formation of earth's climate and aerial spraying of crops. Droplet size plays a crucial role in the dynamics of such processes, and it is therefore desired to measure or estimate it to better understand, predict, or even control this dynamics. State-of-the-art methods for measuring the droplet size distribution use lasers (laser diffraction and phase Doppler particle analysis) or cameras with sizable lenses (image analysis). The interested reader may find a comprehensive review and comparison between the methods in Sijs \textit{et al.} \cite{SizeDistComp2021}. These methods require either expensive equipment, complex setup and calibration, or a combination of the two. The purpose of the present study is to develop a new technique to estimate the size of droplets using inexpensive, off-the-shelf microphones that are able to record the impact of droplets as small as $10\,\micm$ in diameter.  

A condenser microphone is essentially a plates capacitor that records sound by picking up tiny perturbations in the distance between its two parallel plates. Small perturbations in air pressure vibrate one of the plates -- the diaphragm -- and thus alter the voltage on the capacitor, which is then converted into an alteration in pressure. Standard, inexpensive electret microphones can pick sounds as quiet as 30 dB SPL across most of the hearing range (20 Hz -- 20 kHz approximately), making them suitable for very sensitive measurements. Here we exploit the microphone sensitivity by using it to measure pressure perturbations in a liquid, rather than in gas. The dramatic increase in density between a liquid and a gas (in our case a factor of roughly 1000 between the densities of water and air) translates into an increase in pressure, and hence to improved sensitivity compared to air pressure measurements. This was demonstrated by Voermans \cite{mic_measurement_2024} using hydrophones with the goal of measuring the size distribution of ocean spray droplets. Unlike hydrophones, standard microphones cannot operate in a liquid environment, however they can record the impact of small droplets that deposit on their diaphragm and do not damage their mechanism.  

\begin{table}[t!]
    \centering
    \begin{tabular}{c|c|R@{\,--\,}L}
  \textbf{Parameter} & \textbf{Interpretation} & \multicolumn{2}{c}{\textbf{Range}} \\
  \hline
  $R$  & droplet radius         & 1\,\mu\mathrm{m} & 1\,\mathrm{mm} \\
  $u$  & droplet velocity       & 0                & 1\,\mathrm{m/s} \\
  $Bo$ & \textit{Bond} number   & 10^{-7}          & 10^{-1} \\
  $Oh$ & \textit{Ohnesorge} number     & 10^{-3}          & 10^{-1} \\
  $Re$ & \textit{Reynolds} number     & 0                & 10^{3} \\
  $We\equiv Oh^2Re^2$ & \textit{Weber} number & 0      & 10
\end{tabular}

    \caption{Droplet radius and velocity range, given as extreme values, that are considered in the present work, along with the corresponding values of the dimensionless numbers that characterize the impact dynamics. The characteristic values for water droplets surrounded by air -- $\rho=1000\,\rm kg\,m^{-3}$, $\mu=10^{-3}\,\rm Pa\cdot s$, and $\sigma=0.072\,\rm N\,m^{-1}$ -- were used to calculate the dimensionless numbers.}
    \label{params table}
\end{table}

Droplet impact on solid surfaces has been widely investigated in a variety of settings, including liquids of different characteristics and velocities, hydrophilic/hydrophobic and/or hot/cold surfaces, or varying environmental conditions (see Josserand and Thoroddsen's review \cite{impact_review_2016} for recent advances). In the context of water droplets impacting a microphone diaphragm -- typically a flat and smooth aluminum plate surrounded by atmospheric air -- the leading factors affecting the impact dynamics can be classified into three groups: aluminum surface properties, water properties, and droplet kinematic conditions. The first group is characterized by the static aluminum-water contact angle $\alpha$, which was measured to be 70\textdegree-80\textdegree \cite{aluminum_contact_angle_2019}, i.e., well within the hydrophilic range, ans so incoming droplets are very unlikely to bounce off from the surface after impact. The water properties include density ($\rho$), viscosity ($\mu$), and surface tension with the air ($\sigma$), and the droplet kinematics is dictated by its size (radius $R$, assuming droplets are spherical) and its incoming velocity $u$. Those parameters combined, along with the gravitational acceleration $g$, give three dimensionless numbers that help classify the impact regime: the \textit{Bond} number $Bo=\rho g R^2/\sigma$, the \textit{Ohnesorge} number $Oh=\mu/\sqrt{\rho\sigma R}$, and the \textit{Reynolds} number $Re=\rho R u/\mu$. For water droplets in the micro-meter range, i.e., $1\,\rm \mu m<R<1\,\rm mm$, the $Bo$ and $Oh$ numbers are small ($<0.1$) and hence droplets take the form of a spherical cap at equilibrium (due to $Bo\ll1$) and their dynamics in getting to this equilibrium is largely inviscid (due to $Oh\ll1$). Focusing on droplets with low impact velocities, we limit the analysis to $u<1\,\rm m/s$, which is the impact velocity of a free-falling droplet that is released from $h=5\,\rm cm$ with no drag, calculated via $u=\sqrt{2hg}$. In Table \ref{params table} we list the range for $R$ and $u$ along with ranges for $Bo$, $Oh$, and $Re$. A range for the \textit{Weber} number $We=\rho Ru^2/\sigma\equiv Oh^2Re^2$ was added to examine the possibility of droplet splashing, which is typically characterized by a combination of the $Re$ and $We$ numbers. We used the results from Palacios \textit{et al.} \cite{drop_splash_critical_2013} to calculate the critical \textit{Weber} number for the highest $Re$ considered here, $We_c(Re=10^3)$, above which droplet splashing is expected to occur. This critical value of $We_c=2\cdot10^2$ is an order of magnitude larger than the largest $We$ in table \ref{params table}, and hence droplets are not expected to splash.

The dimensionless numbers values suggest that, within the parameter range considered here, droplets are expected to settle into a spherical-cap shape shortly after the initial impact, and oscillate about this shape while decaying to equilibrium. In such case a droplet will, in the late stage of impact, oscillate at distinct resonant frequencies that are dependent on its properties, as well as on the droplet contact-line mobility $\Lambda$, which is a parameter describing how fast the contact line moves in response to a change in the contact angle \cite{Davis1983,contact_line_mobility_measured_2018}. Bostwick and Steen developed a mathematical framework to compute these frequencies \cite{Bostwick2014} and validated the results against experiments \cite{Bostwick2015}, showing excellent overall agreement. Since contact-line mobility can be considered a material property \cite{contact_line_mobility_2022}, it is reasonable to assume that, conditioned on obtaining $\Lambda$ for a prescribed liquid-surface pair, one can measure the resonant frequencies of droplets and invert the process to compute their size.

\begin{figure}
    \centering
    \includegraphics[width=1\linewidth]{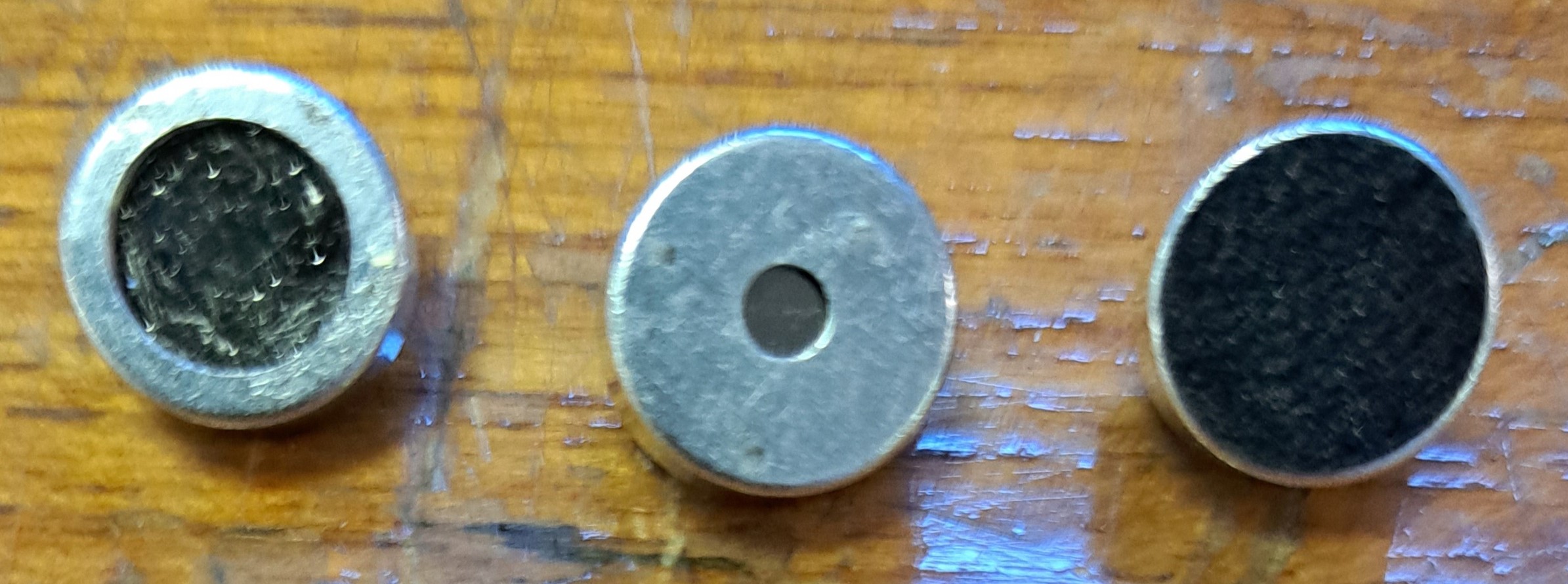}
    \caption{View of electret microphones such as the ones used in the experiments: untouched (right); with its fabric cover peeled off to show the small hole that exposes the diaphragm (center); machined to expose $\sim$50\% of the diaphragm surface area (left).}
    \label{microphone_comparison_pic}
\end{figure}

To test the idea of computing the size of droplets from measurement of their resonant frequencies, we conducted a series of experiments with droplets of varying sizes that impacted an electret condenser microphone diaphragm. Off-the-shelf electret microphones have an aluminum casing that covers the diaphragm and leaves only a small hole, 2 mm in diameter, that allows air pressure perturbations to penetrate through. To increase the exposed microphone surface area, so as to allow for smooth landing of droplets on its diaphragm, we delicately machined the casing without damaging the diaphragm and exposed roughly 50\% of the diaphragm area (see figure \ref{microphone_comparison_pic} for a comparison between standard and machined microphones). We used a pipette (Gilson PIPETMAN P2) to generate droplets of volume $0.30,0.35,\dots,1.05\, \micl$ which, assuming the droplets are spherical, translates to radii in the range 415-630 $\micm$. We repeated the experiment 20 times for each of the 16 droplet volumes, thus recording a total of 320 experiments. 

\begin{figure}
    \centering
    \includegraphics[width=1\linewidth]{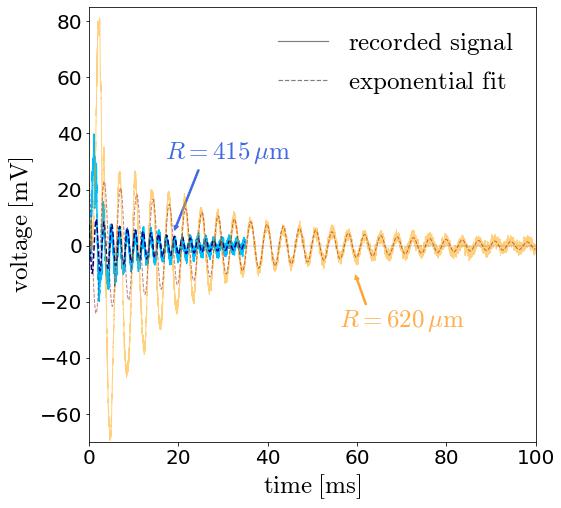}
    \caption{Representative signals of droplets of radii 415 (blue) and 620 $\micm$ (orange) impacting a microphone diagram, showing the recorded voltage as a function of time. The solid and dashed curves are the measurements and the exponential functions fitted to the data, respectively. The signals clearly show the exponentially-decaying nature of the signal at late times, i.e., far enough from the initial impact point, where the two curves (signal and fit) overlap almost entirely.}
    \label{V_vs_t plot}
\end{figure}

\begin{figure}
    \centering
    \includegraphics[width=1\linewidth]{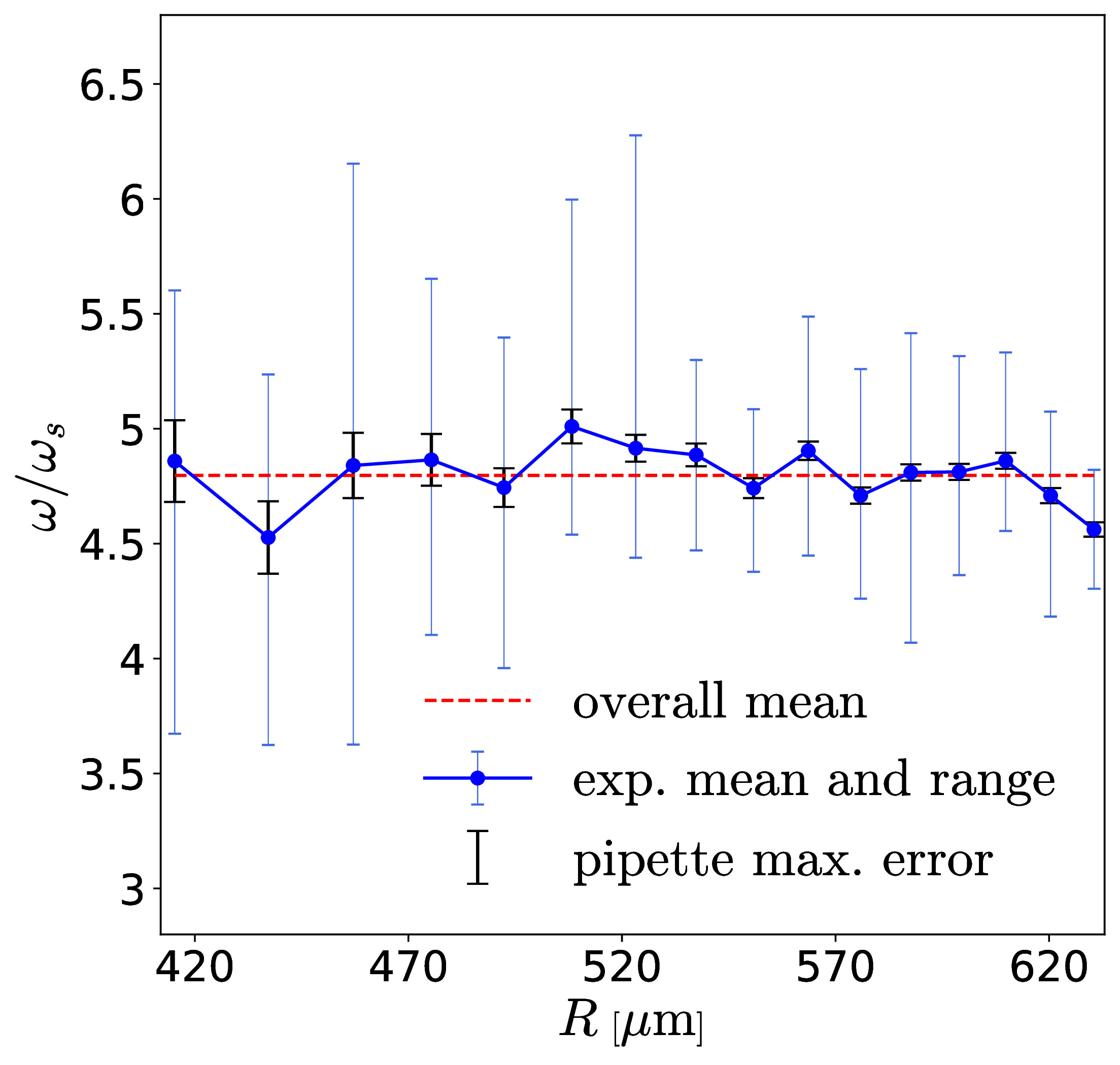}
    \caption{Scaled resonant frequency $\omega/\omega_s$ vs. the droplet radius $R$. The red dashed line marks the mean of $\omega/\omega_s$ over the entire set of experiments; the blue dots and errorbars are the mean and range of $\omega/\omega_s$ over each set of 20 repetitions for each $R$; the black errorbars give the maximum error in estimating $\omega/\omega_s$ using $\omega_s(R)$ given the pipette maximal systematic error for $R$. The measurement range clearly dwarfs the pipette error, which indicates that measurement (and not droplet generation) errors are the main source of uncertainty.}
    \label{omega_vs_R}
\end{figure}

In each experiment the pipette was loaded with water and positioned on a holder aligning it vertically with the microphone center, such that the pipette tip rested 10 mm above the stationary diaphragm. A droplet was then released onto the diaphragm and the impact signal was recorded using an oscilloscope (Rigol DS1054Z) at a sampling rate of 2 MHz, which is 3-4 orders of magnitude faster than the expected droplet oscillation frequencies. Standard oscilloscope triggering was used to capture the droplet impact signal; the threshold at all droplet sizes was set such that background noises as loud as 70 dB did not trigger the capture while the droplet impact did. Following each experiment, we photographed the deposited droplet and extracted its center position $d$ relative to the diaphragm center using image processing. We have set a threshold of $d<1$ mm above which signals were disqualified, and the experiment was repeated. Figure \ref{V_vs_t plot} displays two representative signals (solid curves) of droplets with radii 415 (blue) and 620 $\micm$ (orange). The signals are truncated when the standard deviation is only 5\% above that of the microphone noise. We take the Fourier transform of each signal and extract the first-mode resonant frequency $\omega$ using the maximal Fourier coefficient. We then fit the last 2/3 of each signal, i.e., the late-time stage where we expect the signal to decay exponentially, with a function of the form $\Re\left[a\exp(\ii \Omega t)\right]$, where $a,\Omega\left(\equiv\omega+\ii\gamma\right)\in\mathbb{C}$ such that $\Re\left[a\right]$, $\Im\left[a\right]$, and $\gamma$ are the fitting parameters. The dashed curves in figure \ref{omega_vs_R} are such fitted functions, plotted over the entire signal to show the excellent agreement at late times and the clear mismatch at early times. Each signal is hereinafter characterized by the resonant frequency $\omega+\ii \gamma$.

The expected trend of increase in resonant frequency and decay rate when the droplet size is decreased is easily seen in figure \ref{V_vs_t plot}. According to the traditional scaling of time in inviscid droplets \cite{Bostwick2014} we expect that $\omega\propto\omega_s\equiv\sigma^{1/2}\rho^{-1/2}R_s^{-3/2}$, where
\begin{equation}
R_s=R\left[\left(2+\cos\alpha\right)\csc^{3}\alpha\sin^{4}\left(\alpha/2\right)\right]^{-1/3}
\end{equation}
is the sessile drop spherical cap radius. In figure \ref{omega_vs_R} we plotted the mean of the dimensionless resonant frequency $\omega/\omega_s$ across each set of 20 repetitions (solid, blue curve) along with errorbars that mark the measurement range at each $R$. While the curve is fairly steady about the overall mean, roughly 4.8 marked by the red dashed line, the discrepancies and the large range reveal significant uncertainty in the measurements, which makes the naive estimation of $R=\text{const.}\times\omega^{-2/3}$ impractical. To demonstrate that this large range does not stem from errors in generating consistently-sized droplets, we used the maximal systematic error of the pipette in generating droplets of prescribed volume \cite{Gilson}, and calculated the maximal range for estimating $\omega/\omega_s(R)$. This range is displayed by the black errorbars, which are significantly smaller than the measurement range for all $R$ and thus verify that the large uncertainties are the result of measurement errors. These errors may stem from irregularities in the diaphragm surface, which can lead to variations in the droplet shape that directly affect its resonant frequency, or from variations in the landing position on the diaphragm. In additional experiments included in the supplementary material, we allowed droplets to land anywhere on the surface, given that the entirety of the deposited droplet lies on the diaphragm, and recorded both the impact signal and the distance $d$ from the droplet center to the diaphragm center. These measurements did not show a clear trend of $\omega$ vs. $d$; they did, however, clearly demonstrate the larger variability in $\omega$ as $R$ decreases (see supplementary material). This strengthens our presumption that surface irregularities are the leading cause for the variations in $\omega$: droplet shape should be affected more severely from irregularities of a given size $h$ as $R/h$ decrease, and hence the $\omega$ range measured with a single microphone at varying $R$ is expected to increase with a decrease in $R$. As a result of the large range for $\omega$, we decided to make use of the measured signals to train a neural network, and test whether such network can later predict the droplet size from reading their resonant frequencies.

\begin{figure}
    \centering
    \includegraphics[width=1\linewidth]{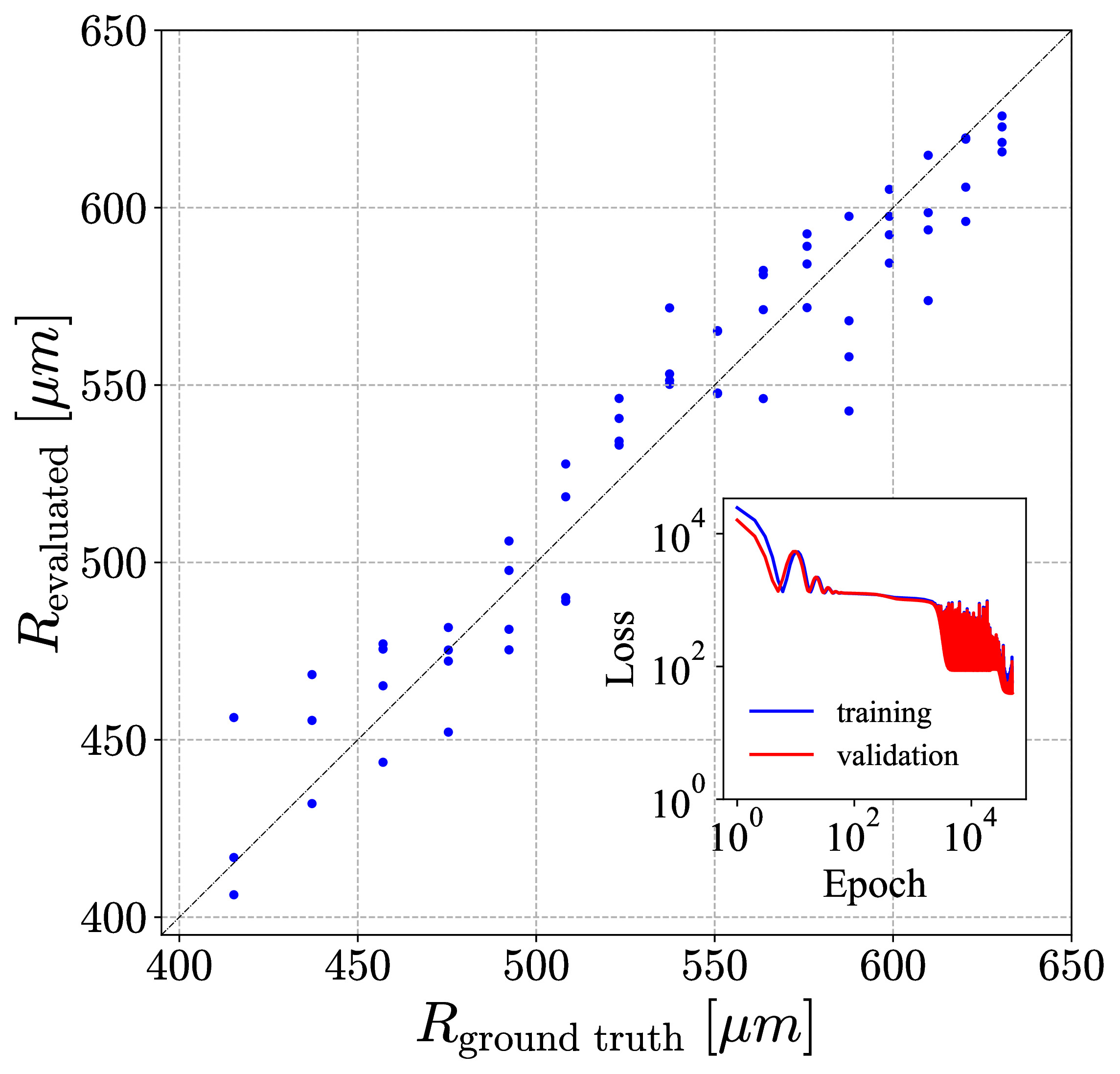}
    \caption{Neural network evaluation of the droplet radii on the test data against the ground-truth results. The dashed line is drawn as visual aid to assess the results. All the results lay relatively close to the line with the largest discrepancy of 42 $\micm$ at $R=587\,\micm$, and the largest relative error of 8.6\% at $R=415\,\micm$. The inset shows the network training convergence, with the blue and red curves marking the training and validation losses, respectively.}
    \label{nn result 1}
\end{figure}

We laid a simple network consisting of two input parameters -- the first-mode resonant frequency components $\omega$ and $\gamma$ fitted to each signal -- and a single output parameter: the droplet radius. We employed a layout with two hidden layers of 20 nodes each, using ReLU as the activation function. Training used mean-square error as the loss function and the Adam optimizer with a learning rate of 0.002, with 20\% of the data reserved for testing. We used the pipette inferred volumes to calculate the droplet radii of spherical droplets, and treated these as ground-truth results. Figure \ref{nn result 1} presents the radii evaluated by the network against their ground-truth counterparts; the dashed diagonal line serves only as aid to visually assess the results. The inset shows the neural network training convergence, where both the training and validation losses follow a similar trend, with noticeable oscillations after $\sim5000$ iterations stemming from the fixed learning rate. All relative errors between the evaluated and ground-truth radii fall below 9\%, and the overall mean is 2.73\%. The relatively small data set used for the network training seems to be the limiting factor in further reducing the error. All our attempts of tweaking the network structure parameters and/or decreasing the learning rate, over which no overfitting was observed, have led to very similar results that were all below 10\% relative error with a mean of 2.5-4\%. 

The results demonstrate the method ability to estimate, with a relative error <10\%, the size of droplets in the tested range from measurement of their impact force using an inexpensive microphone. The range of droplet size used in this work can be extended to smaller and larger droplets. In fact, using the same microphone we were able to record the impact signal of several droplets with $R<100\,\rm \mu m$, which were generated using a micro-goniometer and replicated the same trend shown in figure \ref{V_vs_t plot}. These measurements were not included in the results due to the lack of ground-truth values, however the same trend of decaying oscillations was captured (see supplementary material). Larger droplets can also be measured, however the impact of ones larger than $R=1\,\rm mm$ dramatically increases the probability for mechanical damage to the microphone. This is partly due to the droplet size itself since it occupies more of the diaphragm area, and partly due to an increase in $Bo\propto R^2$ and $Re,We\propto R$ that in turn increases the probability for droplet spreading or splashing on the diaphragm. 

In the present work we measured the impact of droplets one by one and cleared the diaphragm surface before each measurement. Furthermore, to allow for even analysis of the results, we only considered droplets that impact the diaphragm center. In practice, droplets can land on any part of the diaphragm, however, our additional experiments show this does not have a clear effect on the results (see supplementary material); moreover, failure to remove droplets that deposit on the surface can lead to a change in the measured frequency of subsequent droplets. This may occur due to mass added onto the diaphragm that is vibrating, or even due to droplets landing onto deposited ones instead of directly on the diaphragm surface. Since the method can easily be scaled up by stacking multiple microphones and recording the signals in parallel channels, multiple impacts can be recorded simultaneously. The difficulty in removing deposited droplets can be tackled, for example, by stacking a large number of microphones and saving only the first signal that triggers the capture in each one. 

Given the encouraging quantitative agreement in figure $\ref{nn result 1}$ and the prospects for substantial improvement with larger sets of data, the method we developed shows great promise in offering an inexpensive alternative to measuring the size distribution of droplets. Furthermore, the method can operate with background noise and can easily be scaled up. Some technical challenge, such as the prevention of water from penetrating through the microphone mechanism, will need to be resolved before the method can be implemented in real-world settings, however those are beyond the scope of the present work.

\section*{Supplementary material}
See supplementary material for experimental results that were not included in the analysis: impact signals of smaller droplets generated using a microgoniometer, and dependency of $\omega$ on the landing position on the diaphragm.

\begin{acknowledgments}
The author acknowledges the financial support from the Leverhulme Trust in the form of an Early Career Fellowship.
\end{acknowledgments}

\section*{Author Declaration}
    
\subsection*{Conflict of interest}
The author has no conflicts to disclose.

\section*{Data Availability Statement}

The data that support the findings of
this study are openly available in
https://github.com/AVOF/mic-drop.git.


\end{document}